\documentclass[twocolumn,prx,aps,superscriptaddress, longbibliography]{revtex4-1}

\usepackage[utf8]{inputenc}
\usepackage[T1]{fontenc}
\usepackage{graphicx}
\usepackage{dcolumn}
\usepackage{bm}
\usepackage{bbm}
\usepackage{times}
\usepackage[colorlinks,citecolor=blue,linkcolor=red]{hyperref}
\usepackage{color}
\usepackage{xcolor}
\usepackage{comment}
\usepackage{mathrsfs}
\usepackage{amsmath}
\allowdisplaybreaks
\usepackage{amssymb}
\usepackage{braket}
\usepackage{mathtools}
\usepackage{bbold}
\usepackage{appendix}

\usepackage{afterpage}  % Added for control
\usepackage{booktabs}   % three-line table

\hypersetup{
    bookmarks=true,         % show bookmarks bar?
    unicode=false,          % non-Latin characters 
    pdftoolbar=true,        % show Acrobat
    pdfmenubar=true,        % show Acrobat 
    pdffitwindow=false,     % window fit to page when opened
    pdfstartview={FitH},    % fits the width of the page to the window
    pdftitle={},    % title
    pdfauthor={},     % author
    pdfsubject={},   % subject of the document
    pdfcreator={},   % creator of the document
    pdfproducer={}, % producer of the document
    pdfkeywords={} {} {}, % list of keywords
    pdfnewwindow=true,      % links in new window
    colorlinks=true,       % false: boxed links; true: colored links
    linkcolor=blue, %red,          % color of internal links (change box color with linkbordercolor)
    citecolor=blue,        % color of links to bibliography
    filecolor=magenta,      % color of file links
    urlcolor=blue,           % color of external links
    breaklinks=true
}

\begin{document}
\allowdisplaybreaks 
% 设置公式间距

\newcommand{\wrap}[1]{\begin{split}#1\end{split}}
\title{Constructing Exceptional Knots and Links with Arbitrary Braiding Topology}

\author{Bin Jiang}
\affiliation{Department of Modern Physics, School of Physical Sciences, University of Science and Technology of China, Hefei, 230026, China}
\affiliation{Suzhou Institute for Advanced Research, University of Science and Technology of China, Suzhou, 215123, China}

\author{Aolong Guo}
\affiliation{Department of Physics, School of Physical Sciences, University of Science and Technology of China, Hefei, 230026, China}
\affiliation{Suzhou Institute for Advanced Research, University of Science and Technology of China, Suzhou, 215123, China}

%\author{Zhi-Kang Lin}
%\affiliation{New Cornerstone Science Laboratory, Department of Physics, The University of Hong Kong, Hong Kong 999077, China}

\author{Qilin Cai}
\email{qilin_cai@ustc.edu.cn}
\affiliation{Suzhou Institute for Advanced Research, University of Science and Technology of China, Suzhou, 215123, China}
\affiliation{School of Biomedical Engineering, Division of Life Sciences and Medicine, University of Science and Technology of China, Hefei 230026, China}

\author{Jian-Hua Jiang}
\email{jhjiang3@ustc.edu.cn}
\affiliation{State Key Laboratory of Bioinspired  Interfacial Materials Science, Suzhou Institute for Advanced Research, University of Science and Technology of China, Suzhou, 215123, China}
\affiliation{Department of Modern Physics, School of Physical Sciences, University of Science and Technology of China, Hefei, 230026, China}
\affiliation{School of Biomedical Engineering, Division of Life Sciences and Medicine, University of Science and Technology of China, Hefei 230026, China}
\affiliation{School of Physical Science and Technology \& Collaborative Innovation Center of Suzhou Nano Science and Technology, Soochow University, Suzhou 215006, China}

\begin{abstract} 
\textcolor{red}{Exceptional knots and links represent a remarkable class of non-Hermitian metals in which exceptional degeneracies form knotted or linked manifolds in momentum space.}
Here, we report a universal construction framework for realizing exceptional knots and links with arbitrary braiding topology in 3D minimal two-band non-Hermitian systems. 
\textcolor{red}{Our approach combines braid theory with semiholomorphic polynomials to establish a direct correspondence between braid words and non-Hermitian Bloch Hamiltonians.
This framework enables the realization of a broad variety of exceptional configurations, including torus knots, lemniscate knots, nonfibred knots, hyperbolic knots, and multi-component links, within explicit tight-binding Hamiltonians.}
Furthermore, we demonstrate controllable topological transitions in which exceptional knots can be continuously untied through \textcolor{red}{redistribution and reconnection of exceptional points, accompanied by transient exceptional chains and changes in spectral complex energy braiding.} 
Our results establish \textcolor{red}{a universal route toward programmable non-Hermitian knot topology and provide a versatile platform for exploring knotted band degeneracies and their associated physical phenomena across} photonic, acoustic, mechanical, and cold-atom systems.
\end{abstract}

\date{\today}

\maketitle

\section{introduction}

Exceptional degeneracies represent a defining hallmark of non-Hermitian physics, differing fundamentally from their Hermitian counterparts. At their core are exceptional points (EPs)---defective spectral singularities where eigenvalues and eigenvectors coalesce, rendering the Hamiltonian non-diagonalizable~\cite{heiss2004exceptional, EP_2018AIP, EP_2020AIP, RevModPhys_EP}. These degeneracies arise generically in open and driven systems and have been experimentally realized across diverse platforms, including photonics~\cite{NHP_2017Npho, pan2018photonic, miri2019exceptional, li2023exceptional, feng2025non, xiao2025non}, acoustics~\cite{NHA_2021NC, 2023PRL_acoustic, NHA_2024NRP, 2025arXiv_HOER}, mechanics~\cite{2019PRB_machanical, TEC_2023PRL}, electrical circuits~\cite{NHEC_2020PRL, 2025PRL_circuit}, and active matter~\cite{NHAM_2020NC, NHAM_2025ROPP}. Acting as branch points on the complex energy Riemann surface, EPs give rise to striking physical effects such as chiral mode switching~\cite{CMS_2020PRL, CMS_2025NPR}, mode selection~\cite{laha2020exceptional}, nontrivial spectral winding~\cite{SW_2022Nature, SW_2024NC}, enhanced parametric response~\cite{ES_2025LSA, 2025PRL_circuit, 2025PRR_EE} \textcolor{red}{and quantum sensing}~\cite{2025NP_EPsensing} and unidirectional amplification~\cite{2021PRB_amp, 2022CP_amp}, establishing them as fundamental building blocks of non-Hermitian phenomena.

In momentum space, the dimensionality of exceptional degeneracies is governed by codimension arguments: in two-dimensional ($2$D) systems, EPs typically appear as isolated point defects~\cite{2019PRB_isoEP, 2022PRR_nogo, cao2024observation, 2026PRL_singleEP} or in pairs connected by open Fermi arcs~\cite{OFA_2018Science, 2021PRL_FermionDouble, OFA_2025CP, 2015PRB_EPs}, while in three-dimensional ($3$D) systems, they generically extend into one-dimensional ($1$D) manifolds termed exceptional lines. These extended non-Hermitian degeneracies exhibit robustness against perturbations, enabling the formation of diverse configurations including exceptional rings (ERs)~\cite{ER_2017PRL, 2019PRB_machanical, 2025PRL_ER} and high-order ERs~\cite{2025arXiv_HOER}, intersecting exceptional chains (ECs)~\cite{TEC_2023CP, TEC_2023PRL}, and more elaborate geometries~\cite{others_2024PRB}. As embedded curves in $3$D momentum space, exceptional lines possess an intrinsic global geometry that demands a topological characterization beyond mere local stability considerations.

A compelling frontier in non-Hermitian band theory involves engineering nontrivial knot and link topologies within exceptional lines~\cite{knot_2019PRB, knot_2021CP, knot_2021PRL, link_2018PRA, link_2019PRB, link_2020PRB, link_2020PRA}. Knotted structures are fundamental to topology and pervade diverse physical contexts, from fluid vortices~\cite{liquid_2014PRL}, acoustic vortices~\cite{acoustic_2020NC}, and optical fields~\cite{optical_2010NP, wang2025topological, wang2026topological} to polymers~\cite{polymer_2025}, liquid crystals~\cite{liquid_2014PRL}, quantum systems~\cite{quantum_2016NP}, particle physics~\cite{particle_2025PRL}, and topological solitons~\cite{soliton_2017PRL}. Embedding knotted structures into band degeneracies opens new avenues for non-Hermitian topological phases of matter, where abstract knot invariants may acquire direct physical significance through measurable spectral and dynamical signatures. Several prior works have demonstrated exceptional knots and links in specific models~\cite{knot_2019PRB, knot_2021CP, knot_2021PRL, link_2018PRA, link_2019PRB, link_2020PRB, link_2020PRA}, revealing intriguing connections between non-Hermitian topology and exceptional structures.

Despite these advances, a general and constructive principle for realizing exceptional knots with arbitrary braiding topology has remained elusive. Existing approaches typically rely on model-specific designs, implicit fine-tuning, or specialized functional forms, thereby limiting access to only restricted families of knots or links~\cite{knot_2019PRB, lemn_2019SPP}. The fundamental challenge stems from the constrained nature of exceptional degeneracies: exceptional lines emerge as the intersection of \textcolor{red}{two} real conditions in momentum space, and engineering these constraints to form a prescribed knotted curve is highly non-trivial. Moreover, in most cases, knot topology is identified only a posteriori, without a direct correspondence between the desired knot structure and Hamiltonian construction. This hinders systematic exploration and controlled comparison across distinct knot types, leaving open the critical question of whether a universal non-Hermitian framework exists to encode, realize, and continuously transform any desired knot or link at the level of band degeneracies.

In this work, we address these challenges by encoding knot/link information into a pair of scalar functions whose zero-level sets intersect nontrivially.  A key conceptual advance is the integration of braid theory and semiholomorphic polynomials into non-Hermitian band construction. Leveraging Alexander's theorem~\cite{Alexander}, which states that every knot or link can be represented as the closure of a braid, we develop a systematic pipeline: starting from a braid word, we construct a trigonometric braid, promote it to a braid polynomial, and ultimately derive a semiholomorphic polynomial whose nodal set on the three-sphere realizes the target knot or link. By embedding this construction into Bloch momentum space via a suitable mapping from the Brillouin zone to complex coordinates, we obtain tight-binding lattice models that prescribed exceptional knot/link.

The remainder of this work is organized as follows. In Sec.~\ref{constructing}, we develop universal and constructive approach to designing exceptional knots and links, leveraging braid theory and semiholomorphic polynomials to encode arbitrary knot topology into band structures. We then demonstrate the methodology in Sec.~\ref{Examples} through concrete tight-binding models. In Sec.~\ref{untie_TPT}, We further show that exceptional knots can be continuously untied to unknotted or unlinked configurations by increasing the width of braid. In Sec.~\ref{Summary}, we conclude this work with serval outlooks.

\section{Design strategy for exceptional knots} \label{constructing}

An interaction-free two-band non-Hermitian system serves as a minimal yet powerful platform for exploring novel topological phenomena. \textcolor{red}{The traceless Hamiltonian} in momentum space can be expressed in the most general form as \textcolor{red}{
\begin{equation}
    \hat{H}(\boldsymbol{k}) = 
    \left[
    \begin{array}{cc}
        d(\boldsymbol{k}) & b(\boldsymbol{k}) \\
        c(\boldsymbol{k}) & -d(\boldsymbol{k})
    \end{array}
    \right],
    \label{Ham}
\end{equation}
with $d,b,c\in \mathbb{C}$. For convenience in subsequent analysis, we define a complex-valued scaler field
\begin{equation}
    f(\boldsymbol{k}) = d^{2}(\boldsymbol{k})+b(\boldsymbol{k})c(\boldsymbol{k}) \equiv f_1(\boldsymbol{k}) + if_2(\boldsymbol{k}),
    \label{ddbc}
\end{equation}
where $f_1(\boldsymbol{k})$ and $f_2(\boldsymbol{k})$ are real-valued scalar functions. The corresponding energy eigenvalues are given by $E_\pm(\boldsymbol{k}) = \pm\sqrt{d^{2}(\boldsymbol{k})+b(\boldsymbol{k})c(\boldsymbol{k})}\equiv\pm\sqrt{f(\boldsymbol{k})}$.} The condition for spectral degeneracy is $f(\boldsymbol{k})=0$ expect the trivial case $d,b,c=0$, where the Hamiltonian becomes defective, \textcolor{red}{\textit{i.e.}, not only do the eigenvalues coalesce into a single value, but their corresponding eigenvectors (or eigenstates) also unify and become collinear (losing their linear independence).} Decomposing this complex equation into real and imaginary parts yields two fundamental conditions:
\begin{equation}
    \textrm{Re}[E^2] = f_1(\boldsymbol{k}) = 0,~~~~\textrm{Im}[E^2] = f_2(\boldsymbol{k}) = 0.
    \label{E_square}
\end{equation}

In $3$D systems, each constraint in Eq.~(\ref{E_square}) typically defines a $2$D manifold. The intersection of these manifolds generally yields a $1$D closed exceptional line, a continuous locus of  EPs. Exceptional lines are robust topological defects in $3$D momentum space and can give rise to intricate morphology such as unknotted ERs~\cite{ER_2017PRL}, self-intersected ECs~\cite{TEC_2023CP, TEC_2023PRL} as well as exceptional knots~\cite{knot_2019PRB, knot_2021CP, knot_2021PRL} and links~\cite{link_2018PRA, link_2019PRB, link_2020PRB}. These exceptional knots or links are further connected by \textit{Seifert surface}~\cite{knot_2019PRB}. Mathematically, a Seifert surface is defined as an oriented, connected, and open surface whose boundary exactly coincides with the knot or link itself (see Sec.~I of the Supplementary materials). \textcolor{red}{In the physical context, two important realizations of Seifert surfaces emerge~\cite{knot_2019PRB}: the open Fermi surface and the open $i$-Fermi surface. An open Fermi surface is characterized by $\textrm{Re}[E] = 0$, which leads to
\begin{equation}
    \textrm{Re}[E^2] = f_1 \leq 0,~~~~\textrm{Im}[E^2] = f_2 = 0. \label{open_Fermi_surface}
\end{equation}
Similarly, we can define an open $i$-Fermi surface by imposing $\textrm{Im}[E] = 0$ leading to the conditions
\begin{equation}
    \textrm{Re}[E^2] = f_1 \geq 0,~~~~\textrm{Im}[E^2] = f_2 = 0. \label{open_i_Fermi_surface}
\end{equation}
The presence of a finite open or open-$i$ Fermi surface, as described by Eqs.~(\ref{open_Fermi_surface}-\ref{open_i_Fermi_surface}) indicates that these systems exhibit metallic behavior, in contrast to node-line semimetals~\cite{2011PRB_lines, 2017PRB_links} in Hermitian systems, where the density of states vanishes at the Fermi level.} 

\begin{figure}
\centering \includegraphics[width=8.6cm]{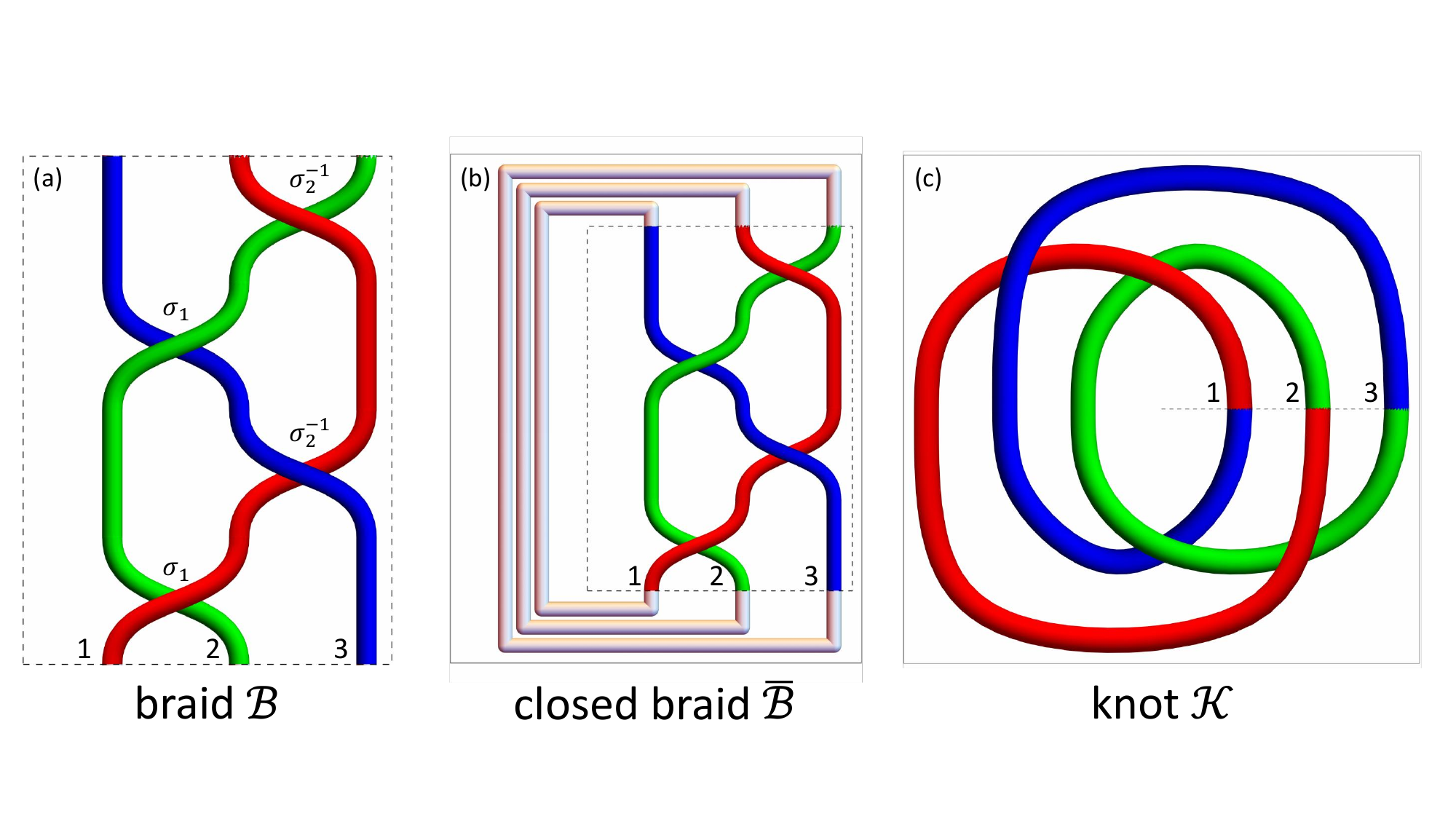}
\caption{\textcolor{red}{Schematic diagram of the correspondence between a braid $\mathcal{B}$ and a knot or link $\mathcal{K}$. Braid $(\sigma_1 \sigma_2^{-1})^2$ and figure-eight knot are illustrated as an example. 
The closed braid $\mathcal{\bar B}$ is formed by closure operation, which connects the respective endpoints on the top and bottom planes without crossing.
The knot $\mathcal{K}$ and the closed braid $\mathcal{\bar B}$ are isotopic and share the same topology.
The braid group generators $\sigma_i (\sigma_i^{-1})$ denotes the $i$-th strand crossing over (under) the $(i+1)$-th strand.}}
\label{Alexander}
\end{figure}

To construct scalar functions $f_1$ and $f_2$ such that the intersection of their kernels generates described knots or links $\mathcal{K}$, a mathematically rigorous method for specifying a knot or link leverages Alexander's theorem~\cite{Alexander}, which asserts that every knot and link $\mathcal{K}$ can be represented as a closed braid $\mathcal{\bar B}$. A braid $\mathcal{B}$ of $s$ strands is defined by $s$ disjoint 3-dimensional trajectories between two parallel planes, without backward motion in the intervening region. \textcolor{red}{A $2$D projection called braid diagram is usually used to represent the braid as shown in Fig.~\ref{Alexander}(a).} The crossings in the braid diagram correspond to the braid group generators $\sigma_i^{\pm 1}$, where $\sigma_i$ denotes the $i$-th strand crossing over the $(i+1)$-th strand, and $\sigma_i^{-1}$ denotes the under-crossing. The closed braid $\mathcal{\bar B}$ is formed by an operation known as closure, which connects the respective endpoints on the top and bottom planes in a one-to-one fashion without crossing \textcolor{red}{as shown in Fig.~\ref{Alexander}(b)}. The topology of a knot or link $\mathcal{K}$ isotopic to $\mathcal{\bar B}$ is entirely determined by the sequence of crossings in the braid diagram, and algebraically encoded as a braid word, a sequence of braid group generators. The braid group generators obey two fundamental relations: the commuting relation between non-adjacent strands $\sigma_i\sigma_j = \sigma_j\sigma_i$ for $|i-j| \geq 2$, and the braid relation $\sigma_i\sigma_{i+1}\sigma_i = \sigma_{i+1}\sigma_i\sigma_{i+1}$, which underpins the Yang-Baxter equation~\cite{knotBook}. It is crucial to note that due to these relations and Markov moves, the mapping from a braid word to a knot type is many-to-one. Nevertheless, any specific braid word uniquely defines a knot, providing an unambiguous starting point for constructing a Hamiltonian. For instance, the braid word $\sigma_1^2$ yields the Hopf link, $\sigma_1^3$ produces the Trefoil knot, \textcolor{red}{and $(\sigma_1 \sigma_2^{-1})^2$ produces the figure-eight knot, which is plotted in Fig.~\ref{Alexander}. Additionally, reference~\cite{gittings2004minimum} also presents an optional algebraic method for finding the $minimum~braids$ of all knots and links, which have the fewest number of braid crossings and strands (although not unique).}

%%%%%%%%%%%%%%%%%%%%%%%%%%%%%%%%%%%
%%%%%%%%%%%%%%%%%%%%%%%%%%%%%%%%%%%
%%%%%%%%%%%%%%%%%%%%%%%%%%%%%%%%%%%
%%%%%%%%%%%%%%%%%%%%%%%%%%%%%%%%%%%

By virtue of the completeness of the Fourier basis, a braid $\mathcal{B}$ on $s$ strands is parametrized as
\begin{equation}
    B_\mathcal{K} = \bigcup^{s-1}_{j=0}\big(X(t + 2\pi j), Y(t + 2\pi j), t\big),~~t\in[0, 2\pi),
\end{equation}
where $X(t)$ and $Y(t)$ are trigonometric functions. The function $X(t)$ controls the transverse positions of the strands, generating intersections at parameters $t_i$ corresponding to crossings, while $Y(t)$ determines the signs (positive for over-crossings, negative for under-crossings) based on the strand order and knot topology. 
%Crucially, the choice of $X(t)$ and $Y(t)$ appears to have significant functional freedom: multiple distinct trigonometric expressions can reproduce the same knot topology, provided they satisfy the required crossing sequence and strand ordering. 
\textcolor{red}{Given a braid representation, the functions X(t) and Y(t) can be systematically obtained through finite Fourier interpolation of strand trajectories, following Refs.~\cite{methods_2019JOKT, methods_2021PRL, knot52_2017JPA}. 
Although the interpolation is not unique, different choices preserve the braid topology and therefore lead to the same knot/link type after closure. 
The trigonometric interpolation method from~\cite{methods_2019JOKT, methods_2021PRL} is algorithmic but may produce complex parameterizations.
In our work}, the functions $X(t)$ and $Y(t)$ are typically chosen as finite Fourier series in the sense of fewer terms and favoring rational coefficients and frequencies, such as $X(t) = \sum_{k} a_k \cos(\omega_k t + \phi_k), \quad Y(t) = \sum_{k} b_k \sin(\gamma_k t + \psi_k)$, where the coefficients $a_k$, $b_k$, frequencies $\omega_k$, $\gamma_k$, and phase $\phi_k$ and $\psi_k$ are carefully selected to replicate the crossing sequence of the braid word. 
\textcolor{red}{For example, the parameterizations for knots up to six crossings only have two terms for $X$ and $Y$ each (see the following Sec.~\ref{Examples}).} 

To construct a complex-valued polynomial function whose nodal set is the desired knot or link $\mathcal{K}$, we proceed by encoding the trigonometric braid representation into a polynomial map. We define the braid polynomial $\rho_a(u, t)$ as a function of a complex variable $u$ and the braid parameter $t$,
\begin{equation}
    \rho_a(u, t) = \prod^{s-1}_{j=0}\big(u - a[X(t + 2\pi j) + iY(t + 2\pi j)]\big), 
\end{equation}
where $a>0$ is a scaling parameter that controls the geometric width of the braid trajectory. For sufficiently small $a$, the braid trajectory lies close to the origin in the $u$-plane, ensuring that the nodal set on $\mathcal{S}^3$ is isotopic to the closed braid~\cite{knot52_2017JPA, methods_2019JOKT}, \textit{i.e.}, the desired knot or link $\mathcal{K}$. This polynomial is of degree $s$ in $u$, and its roots trace the positions of the braid strands in the complex $u$-plane. To close the braid and embed it into a higher-dimensional space, we generalize $\rho_a(u, t)$ to a semi-holomorphic function $F_a(u, v, \bar{v})$ by replacing the dependence on $t$ with the complex variable $v$. Specifically, we set $t=arg(v)$ and use de Moivre's theorem to express $cos(nt)\rightarrow\frac{v^n+\bar{v}^n}{2}$ and $sin(nt)\rightarrow\frac{v^n-\bar{v}^n}{2i}$. This yields a semi-holomorphic polynomial (holomorphic in $u$ but depending on both $v$ and $\bar{v}$) which can be written as
\begin{equation}
    F_a(u, v, \bar{v}) = \mathrm{polynomial~in}~u,v,\bar{v}.
\end{equation}

The function $F_a$ is defined on $\mathbb{C}^2$, but the key challenge lies in mapping from Brillouin zone $\mathbb{T}^3$ to $\mathbb{C}^2$ while preserving the intricate exceptional knot. To ensure that $f$ encapsulates the braid information, we decompose it into a composition of mappings~\cite{commapping_2022NC}
\begin{equation}
    f = F_a\circ G:~\mathbb{T}^{3} \xrightarrow{G} \mathbb{C}^{2} \xrightarrow{F_a} \mathbb{C}, \label{com_mapping2}
\end{equation}
where $G$ is a parametrization mapping from the Brillouin zone $\mathbb{T}^3$ to $\mathbb{C}^2$, designed to preserve Bloch periodicity using trigonometric functions, 
\begin{equation}
    \begin{split}
        &v(\boldsymbol{k}) = sin(k_x) + isin(k_y),\\
        &u(\boldsymbol{k}) = P\sum_{\alpha=x,y,z} cos(k_\alpha) - Q +isin(k_z),
    \end{split}
\end{equation}
where \textcolor{red}{the real parameters $P$ and $Q$ control the spatial extent of the exceptional knots in momentum space. For all examples considered in this work, they are chosen such that the resulting exceptional knots and links are fully contained within the Brillouin zone. As a result, periodic boundary conditions do not introduce additional isolated EPs or boundary-induced reconnections, and the knot topology is faithfully preserved under the mapping $G$. Finally, Equation~(\ref{ddbc}) always admits a set of solutions
\begin{equation}
    d=0,~~b=1,~~c(\boldsymbol{k}) = f(\boldsymbol{k})=f_1(\boldsymbol{k}) + if_2(\boldsymbol{k}).
    \label{trivial_solution}
\end{equation}
The corresponding tight-binding model is then obtained by inserting Eq.~(\ref{trivial_solution}) into Eq.~(\ref{Ham}).} The procedure for constructing \textcolor{red}{exceptional knots with arbitrary braiding braiding topology in two-band non-Hermitian model} is systematic and general, as summarized by the following theorem:

\textbf{Theorem 1.} For every knot or link $\mathcal{K}$, (1) there is a function $F_a: \mathbb{C}^2 \rightarrow \mathbb{C}$ such that $F_a^{-1}(0)\cap \mathcal{S}^3 = \mathcal{K}$ and $F_a$ is a polynomial in complex variables $u$, $v$ and $\bar{v}$, \textit{i.e.}, $F_a(u, v, \bar{v})$ is a semiholomorphic polynomial. \textcolor{red}{(2) there is a periodic function $f: \mathbb{R}^3 \rightarrow \mathbb{C}$ such that $f^{-1}(0)\cap \mathbb{T}^3 = \mathcal{K}$. (3) The maximum range of hopping in the tight-binding model obtained by inserting Eq.~(\ref{trivial_solution}) into Eq.~(\ref{Ham}) is determined by the degree of the semiholomorphic polynomial $F_a(u, v, \bar{v})$, defined as the greatest sum of exponents of $u$, $v$ and $\bar{v}$ in any single term.}

%%%%%%%%%%%%%%%%%%%%%%%%%%%%%%%%%%%
%%%%%%%%%%%%%%%%%%%%%%%%%%%%%%%%%%%
\begin{figure*}
\begin{center}
\centering \includegraphics[width=17.8cm]{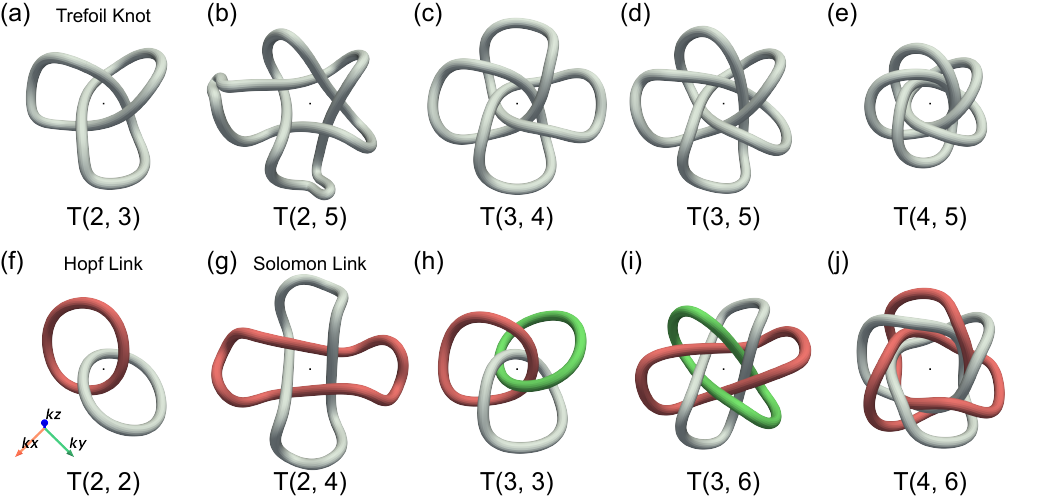}
\caption{Exceptional knots and links on torus $T(p, q)$ in tight-binding models. The fundamental distinction between torus knots and links is determined by the greatest common divisor $r = gcd(p,q)$. (a-e) Torus knots if $r=1$. (f-j) Torus links if $r>1$. The values of $p$ and $q$ are given at the bottom of each panel. $(P,Q)=(1,2.5)$ is adopted for all panels except panels (e) and (j) with $(P, Q)=(2,5)$. \textcolor{red}{The small squares in each panel indicate the center of Brillouin zone ($\Gamma=(0,0,0)$ point).}}
\label{Torus_TBM}
\end{center}
\end{figure*}
%%%%%%%%%%%%%%%%%%%%%%%%%%%%%%%%%%%
%%%%%%%%%%%%%%%%%%%%%%%%%%%%%%%%%%%

\section{Examples in tight-binding models} \label{Examples}

We have established a universal theoretical for constructing exceptional knots in above Sec.~\ref{constructing} and now detailed its application to torus knots/links (Sec.~\ref{subsec_Torus}) and lemniscate knots (Sec.~\ref{subsec_lemniscate}). To demonstrate the scope and utility of this constructing beyond these specialized families, we now apply it to other knots and links (Sec.~\ref{subsec_others}), presented in order of increasing complexity. The sequence then progresses to the Three-twist knot $5_2$ (the simplest non-fibered knot, and the Miller-Institute knot $6_2$ (fibred and hyperbolic knot). Finally, we generalize to multi-component exceptional links such as the Whitehead link $5_1^2$, illustrating the framework's seamless extension to links.

\subsection{Torus knots and links} \label{subsec_Torus}

Torus knots and links, denoted as $T(p,q)$, are closed cycles embedded on the surface of a standard torus, where $p$ and $q$  are integers representing the number of longitudinal and meridional winding cycles, respectively. The fundamental distinction between these topological objects is determined by the greatest common divisor $r = gcd(p,q)$. If $r = 1$, the configuration forms a torus knot, a single, embedded curve, whereas if $r > 1$, the parameters $p$ and $q$ are not coprime, resulting in a torus link, a collection of $r$ mutually interlinked, simpler torus knots $T(p/r, q/r)$. 

A standard braid representation for $T(p,q)$ is given by the closure of a braid on $p$ strands with the braid word $(\sigma_1\sigma_2...\sigma_{p-1})^q$. This braid word describes a sequence where the $p$ strands undergo a full set of adjacent crossings, repeated $q$ times, before being connected to form the closed knot or link. For a braid on $p$ strands, the parametrization is
\begin{equation} 
B_{T(p, q)} = \bigcup_{j=0}^{p-1} \big(X_T(qt+2\pi j), Y_T(qt+2\pi j), t\big), t\in[0, 2\pi), 
\end{equation}
where the functions controlling the transverse and longitudinal positions are elegantly given by
\begin{equation}
X_T(t) = cos(\frac{t}{p}),~~~~Y_T(t) = sin(\frac{t}{p}).
\end{equation}
This parametrization generates a braid where each strand traces a circular path in the transverse plane, with a phase shift of $2\pi/p$ between adjacent strands. The resulting closed braid $\bar{B}_{T(p, q)}$ is isotopic to the torus knot/link $T(p,q)$. 

In fact, for torus knots/links, the entire construction yields a polynomial that is holomorphic (indeed independent of $\bar{v}$), given by 
\begin{equation}
F^{pq}(u, v) = u^p - v^q, \label{FTorus}
\end{equation}
\textcolor{red}{which reproduces the result in~\cite{knot_2019PRB}.} Figure~\ref{Torus_TBM} presents a series of representative torus exceptional knots/links and their open fermi surfaces for various ($p$, $q$) pairs. When the greatest common divisor is $r=gcd(p, q)=1$, the non-Hermitian system hosts a single exceptional knot, such as the familiar Trefoil knot $T(2,3)$ illustrated in Fig.~\ref{Torus_TBM}(a). By contrast, when $r>1$ the system supports an $r$-component torus exceptional link, such as the two-component Hopf link $T(2,2)$ and Solomon link $T(2,4)$, displayed in Figs.~\ref{Torus_TBM}(f) and (g) respectively. 

\textcolor{red}{We note that if either $p$ or $q$ is even, the maximum hopping range in the tight-binding model is halved relative to the generic construction given by Eq.~(\ref{trivial_solution}). Consequently, the Hamiltonian takes the form
\begin{equation}
    \begin{cases}
        d = u^{p/2}, b = -v^{\lfloor q/2 \rfloor}, c = v^{\lceil q/2 \rceil} ~~~~ \textrm{if}~p~ \textrm{is even}, \\    
        d = v^{q/2}, b = -u^{\lfloor p/2 \rfloor}, c = u^{\lceil p/2 \rceil} ~~~~ \textrm{if}~q~ \textrm{is even},
    \end{cases}
    \label{torus_solution}
\end{equation}
where $\lfloor \bullet \rfloor(\lceil \bullet \rceil)$ denotes the floor(ceiling) function. The second line of Eq.~(\ref{torus_solution}) holds because torus knots/links satisfy the symmetry $T(p,q) = T(q,p)$, so the construction for even $q$ is simply obtained by swapping $p$ and $q$ in the even-$p$ case.}

\subsection{Lemniscate knots and links} \label{subsec_lemniscate}

Lemniscate knots~\cite{lemn_2019SPP} generalize the concept of torus knots through their construction as closures of braids derived from generalized lemniscate curves. They are systematically classified by three integers $L(m, n, l)$, where $m$ is the number of strands in the braid, $n$ as the number of rotational periods, and $l$ the lemniscate exponent governs the transverse complexity. This classification establishes a natural hierarchy where the case $l = 1$ is topologically equivalent to the standard torus knot $T(m, n)$, while $ l \geq 2$ yields the distinctive hypocycloidal geometry characterizing non-torus lemniscate knots.

The braid word for lemniscate knots can be directly read off from trigonometric parameterization through a systematic analysis of the strand crossings. The braid representation for lemniscate knots follows the general parameterization
\begin{equation} 
B_{L(m, n, l)} = \bigcup_{j=0}^{m-1} \big(X_L(nt+2\pi j), Y_L(nt+2\pi j), t\big), t\in[0, 2\pi),
\label{B_lemniscate}
\end{equation}
with transverse and longitudinal coordinate functions~\cite{knot52_2017JPA}
\begin{equation}
X_L(t) = cos(\frac{t}{m}),~~~~Y_L(t) = \frac{1}{l}sin(\frac{lt}{m}).
\label{XY_lemniscate}
\end{equation}
The braid word is read off by analyzing the strand crossings in the $(t, X)$-plane projection. Crossings occur at parameter values $t_c$ where the $X$-coordinates of two distinct strands coincide, \textit{i.e.}, $X_L(nt_c+2\pi j) = X_L(nt_c+2\pi k)$ for $j \neq k$. The sign of each crossing (over or under) is determined by comparing the corresponding $Y$-coordinates at $t_c$: if $Y_L(nt_c+2\pi j) > Y_L(nt_c+2\pi k)$, stand $j$ crosses over strand $k$, corresponding to a positive generator $\sigma_i$; otherwise, it corresponds to $\sigma_i^{-1}$. 

The parameter $l$ plays a crucial role in modulating the crossing sequence. For $l=1$, the functions $X_L(t)$ and $Y_L(t)$ describe perfect circles, and the braid word reduces to the regular pattern $(\sigma_1\sigma_2...\sigma_{m-1})^n$, which is the standard braid word for the torus knot $T(m,n)$. For $l\geq2$, the lemniscate exponent $l$ in $Y_L(t)$ compresses the sinusoidal motion in the $Y$-direction, leading to a higher density of crossings and a more complex sequence of over- and under-crossings. The crossings associated with odd-indexed generators (\textit{e.g.}, $\sigma_1,\sigma_3,...$) typically occur at parameter values $t_c=2k\pi/m$ for $k=0, 1,..,m-1$, while the crossings for even-indexed generators (\textit{e.g.}, $\sigma_2,\sigma_4,...$) occur at $t_c=(2k+1)\pi/m$. The resulting braid word becomes $\mathcal{B}_{L(m,n,l)}=(\sigma_1^{\epsilon_1} \sigma_3^{\epsilon_3} ... \sigma_2^{\epsilon_2} \sigma_4^{\epsilon_4}...)^n$, where the exponents $\epsilon_i\in\left\{ +1,-1 \right \}$ are determined by the relative $Y$-coordinates at the crossing points. The sequence of $\epsilon_i$ may exhibit periodic patterns or symmetries reflecting the hypocycloidal nature of the lemniscate curve. For example, when $l=2$, the braid word often exhibits an alternating pattern of signs, while for larger $l$ the pattern can become more intricate, involving longer cycles of over- and under-crossings.

\begin{figure}
\begin{center}
\centering \includegraphics[width=8.6cm]{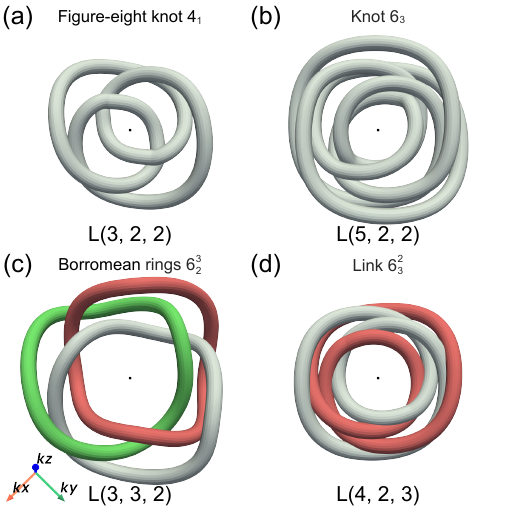}
\caption{Exceptional knots/links on lemniscate $L(m, n, l)$ in tight-binding models. The values of $m$, $n$ and $l$ (Their names in Rolfsen Knot Table) are given at the bottom (top) of each panel. (a) The Figure-eight exceptional knot $4_1$ is obtained from Eq.~(\ref{Fa_L3n2}) with $(n, a, P, Q) = (2, 1, 2, 5)$. (b) The exceptional knot $6_3$ is obtained from Eq.~(\ref{Fa_L5n2}) with $(n, a, P, Q) = (2, 1, 2, 4.6)$. (c) The Borromean exceptional rings $4_1$ is obtained from Eq.~(\ref{Fa_L3n2}) with $(n, a, P, Q) = (3, 1, 2, 4)$. (d) The exceptional link $6_3^2$ is obtained from Eq.~(\ref{Fa_L4n3}) with $(n, a, P, Q) = (2, 1, 2, 5)$. \textcolor{red}{The small squares in each panel indicate the center of Brillouin zone ($\Gamma=(0,0,0)$ point).}}
\label{Lemn_TBM}
\end{center}
\end{figure}

Applying the general procedure outlined in Sec.~\ref{constructing}, \textcolor{red}{for the family $L(3,n,2)$ whose braid word is $(\sigma_1\sigma_2^{-1})^n$}, the corresponding semiholomorphic polynomial is derived as
\begin{equation}
\begin{split}
    F_a(u, v, \bar{v}) &= 64u^3-12a^2u[3+2(v^n-\bar{v}^n)] \\
    &\quad -a^3[14(v^n+\bar{v}^n)+(v^{2n}-\bar{v}^{2n})].
    \label{Fa_L3n2}
\end{split}
\end{equation}
\textcolor{red}{In particular, the case $n=2$($n=3$) corresponds to the Figure-eight knot(Borromean rings), as shown in Figs.~\ref{Lemn_TBM}(a) and (c). Similarly, for the family $L(5,n,2)$}, which involves a $5$-strand braid with word $(\sigma_1\sigma_3^{-1}\sigma_2\sigma_4^{-1})^n$, the polynomial is given by
\begin{equation}
\begin{split}
    F_a(u, v, \bar{v}) &= 1024u^5 - 960a^2u^3 -160a^3u^2(v^2 \\
    &\quad + \bar{v}^2) + 20a^4u[21 - 10(v^n - \bar{v}^n)] \\
    &\quad - a^5[82(v^n + \bar{v}^n) + (v^{2n} - \bar{v}^{2n})]. 
    \label{Fa_L5n2}
\end{split}
\end{equation}
\textcolor{red}{$n=2$ yields the $6_3$ knot at $(a, P, Q) = (1, 2, 4.6)$, as shown in Fig.~\ref{Lemn_TBM}(b). For the simplest non-alternating family $L(4,n,3)$ with braid word $(\sigma_2\sigma_1^{-1}\sigma_3^{-1})^n$, the polynomial reads
\begin{equation}
    \begin{split}
        F_a(u, v, \bar{v}) &= 20736u^4-576a^2u^2[8+3(v^n - \bar{v}^n)]\\
        &\quad +a^4(92-39v^n-231v^n+6v^{2n} \\
        &\quad +30\bar{v}^{2n} -v^{3n}-\bar{v}^{3n}).
        \label{Fa_L4n3}
    \end{split}
\end{equation}
For instance, the two-component exceptional link $L(4,2,3)$ at $(a, P, Q) = (1, 2, 5)$ is plotted in Fig. ~\ref{Lemn_TBM}(d). Eqs.~(\ref{Fa_L3n2}-\ref{Fa_L4n3}) are consistent with the result in~\cite{bode2017knotted}.}

%%%%%%%%%%%%%%%%%%%%%%%%%%%%%%%%%%% 
%%%%%%%%%%%%%%%%%%%%%%%%%%%%%%%%%%%
%%%%%%%%%%%%%%%%%%%%%%%%%%%%%%%%%%%
%%%%%%%%%%%%%%%%%%%%%%%%%%%%%%%%%%%

%%%%%%%%%%%%%%%%%%%%%%%%%%%%%%%%%%% 
%%%%%%%%%%%%%%%%%%%%%%%%%%%%%%%%%%%
\begin{figure}
\begin{center}
\centering \includegraphics[width=8.6cm]{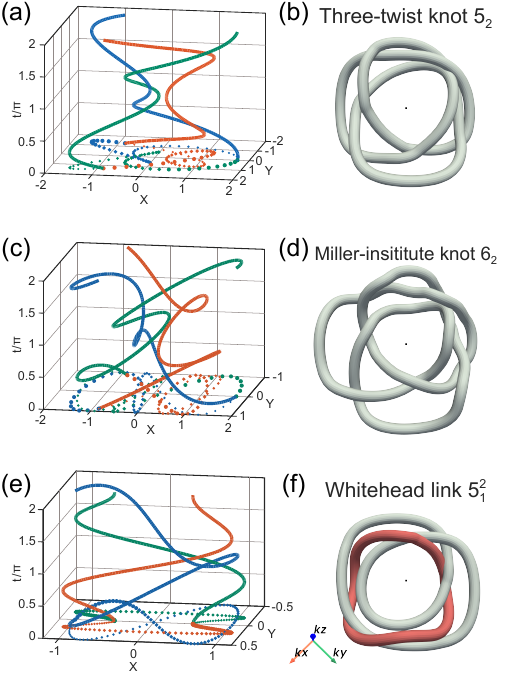}
\caption{Other non-torus exceptional knots in tight-binding model. (a) Trigonometric braid of the knot $5_2$ (solid lines) and its projection on $XY$-plane (dotted lines) are obtained from Eq.~(\ref{XY_K52}). (b) The Three-twist exceptional knot $5_2$ is obtained from Eq.~(S4) with $(a, P, Q) = (0.25, 1, 2.1)$. (c) Trigonometric braid of the knot $6_2$ (solid lines) and its projection on $XY$-plane (dotted lines) are obtained from Eq.~(\ref{XY_K62}). (d) The Miller-insititute exceptional knot $6_2$ is obtained from Eq.~(S8) with $(a, P, Q) = (0.25, 1, 2.1)$. (e) Trigonometric braid of the link $6_2^3$ (solid lines) and its projection on $XY$-plane (dotted lines) are obtained from Eq.~(\ref{XY_Whitehead}). (f) The Whitehead exceptional link $5_1^2$ is obtained from Eq.~(S12) with $(a, P, Q) = (0.25, 2, 4)$. \textcolor{red}{The small squares in panels (b, d, f) indicate the center of Brillouin zone ($\Gamma=(0,0,0)$ point).}}
\label{other_TBM}
\end{center}
\end{figure}
%%%%%%%%%%%%%%%%%%%%%%%%%%%%%%%%%%%
%%%%%%%%%%%%%%%%%%%%%%%%%%%%%%%%%%%

%%%%%%%%%%%%%%%%%%%%%%%%%%%%%%%%%%% 
%%%%%%%%%%%%%%%%%%%%%%%%%%%%%%%%%%%
\begin{figure}
\begin{center}
\centering \includegraphics[width=8.6cm]{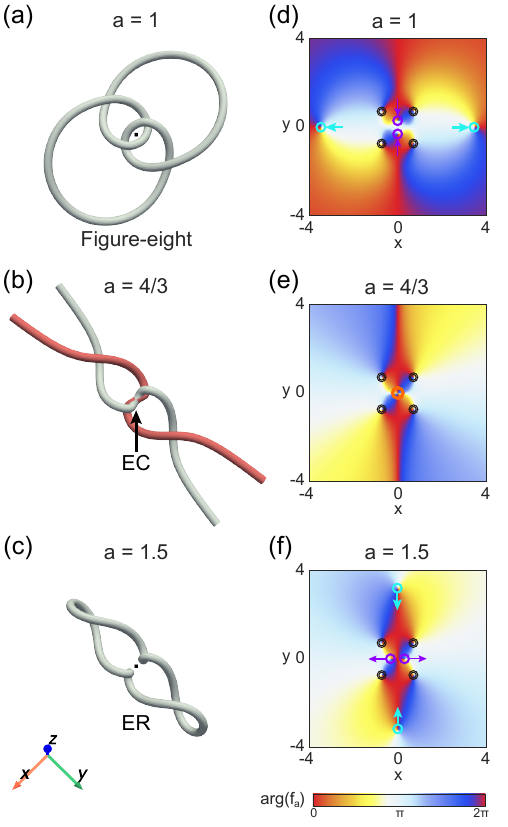}
\caption{The exceptional lines obtained from Eq.~(\ref{Fa_L3n2}) and their evolutions by continuously increasing scaling parameter $a$. (a-c) Evolution of the morphology of exceptional knots (main panels) and their open Fermi surface (insets) by continuously increasing the scaling parameter $a$, starting from exceptional knot $4_1$ with $a=1$ in Fig.~\ref{K41_PT}(a). The exceptional lines are labeled according to the context of the main text. (d-f) Color scale: $arg(f_a)$ in $xy$-plane. Circles: vortices in $arg(f)$ by using standard stereographic mapping Eq.~(\ref{stereographic_coordinates}), which correspondence to exceptional points (EPs). The arrows indicate the moving of EPs. The value of scaling parameter $a$ is given at the top of each panel.}
\label{K41_PT}
\end{center}
\end{figure}
%%%%%%%%%%%%%%%%%%%%%%%%%%%%%%%%%%%
%%%%%%%%%%%%%%%%%%%%%%%%%%%%%%%%%%%

\subsection{Other knots and links} \label{subsec_others}

\textcolor{red}{Not every knot is a torus knot, nor does every braid possess symmetry. Finding a suitable trigonometric braid for a given knot is a crucial step in constructing a Hamiltonian with an exceptional knot or link. In this subsection, we give some examples on how to choose trigonometric functions $X(t)$ and $Y(t)$.}

The Three-twist knot $5_2$ is distinguished as the simplest non-fibred knot. The minimal braid word for knot $5_2$ is $\sigma_1^{-1}\sigma_2\sigma_1^{3}\sigma_2$, which encodes the full sequence of crossings for a $3$-strand braid. To construct a parametric representation that captures the crossing structure of this braid, \textcolor{red}{we employ a modulation strategy: a lower-frequency carrier wave is modulated by a higher-frequency envelope.
Specifically, the transverse coordinate $X$ is constructed such that spatial frequencies of $2/3$ and $5/3$ are introduced, with phase interventions occurring at $t=\pi$ and $t=5\pi$. The longitudinal coordinate $Y$ is chosen analogously. Following this approach, we adopt the parameterization from}~\cite{knot52_2017JPA}
\begin{equation}
    \begin{split}
        X(t) = cos(\frac{2t}{3})+\frac{3}{4}cos(\frac{5t}{3}),\\
        Y(t) = -sin(\frac{4t}{3})-\frac{1}{2}sin(\frac{t}{3}). \label{XY_K52}
    \end{split}
\end{equation}
\textcolor{red}{Applying the aforementioned procedure yields the exceptional knot $5_2$ at $(a, P, Q) = (0.25, 1, 2.1)$, as depicted in Fig.~\ref{other_TBM}(b).}

For the Miller-Institute knot $6_2$, the minimal braiding word is $\sigma_1\sigma_2^{-1}\sigma_1^{3}\sigma_2^{-1}$. Notably, it shares the same unsigned braid word as the knot $5_2$, meaning the sequences of crossing locations are identical. However, the signs (over/under assignments) of the crossings differ between the two knots, leading to distinct knot types. Therefore, we reuse the same $X(t)$ function to replicate the shared crossing sequence, but must design a new $Y(t)$ function to generate the specific sign pattern encoded in the $6_2$ braid word, so~\cite{knot52_2017JPA}
\begin{equation}
X(t) = cos(\frac{2t}{3})+\frac{3}{4}cos(\frac{5t}{3}),~~~~Y(t) = sin(\frac{7}{3}t).
\label{XY_K62}
\end{equation}
\textcolor{red}{The corresponding exceptional knot $6_2$ at $(a, P, Q) = (0.25, 1, 2.1)$ is shown in Fig.~\ref{other_TBM}(d).}

The construction of exceptional links follows precisely the same methodological framework established for exceptional knots. The key distinction is that a exceptional link comprises multiple components, each of which corresponds to a specific subset of strands in its braid representation. For the Whitehead link, its minimal braid has three strands and five crossings, with the braid word $\sigma_1\sigma_2^{-1}\sigma_1\sigma_2^{-1}\sigma_1$. In this representation, strands $2$ and $3$ together form one component of the link, which traces a simple loop. The remaining strand (strand $1$) constitutes the second component, weaving around the first in a lemniscate-like trajectory [see Fig.~\ref{other_TBM}(e)]. This braid structure is captured by the following trigonometric parametrization, where the two components are naturally distinguished by their periods~\cite{knot52_2017JPA}
\begin{equation}
    \begin{split}
        X_1(t) = \frac{1}{2}cos(\frac{3}{2}t)-cos(\frac{1}{2}t),~~Y_1(t) = \frac{1}{4}sin(\frac{1}{2}t); \\
        X_2(t) = -cos(t),~~Y_2(t) = \frac{1}{2}sin(2t).
        \label{XY_Whitehead}
    \end{split}
\end{equation}
\textcolor{red}{The exceptional Whitehead link $5_1^2$ at $(a, P, Q) = (0.25, 2, 4)$ is shown in Fig.~\ref{other_TBM}(f). Details of the calculations for semi-holomorphic polynomials $F_a(u,v,\bar{v})$ and the Hamiltonian are provided in Secs.~III-VI of the Supplementary materials.}

%%%%%%%%%%%%%%%%%%%%%%%%%%%%%%%%%%% 
%%%%%%%%%%%%%%%%%%%%%%%%%%%%%%%%%%%
\begin{figure*}
\begin{center}
\centering \includegraphics[width=17.8cm]{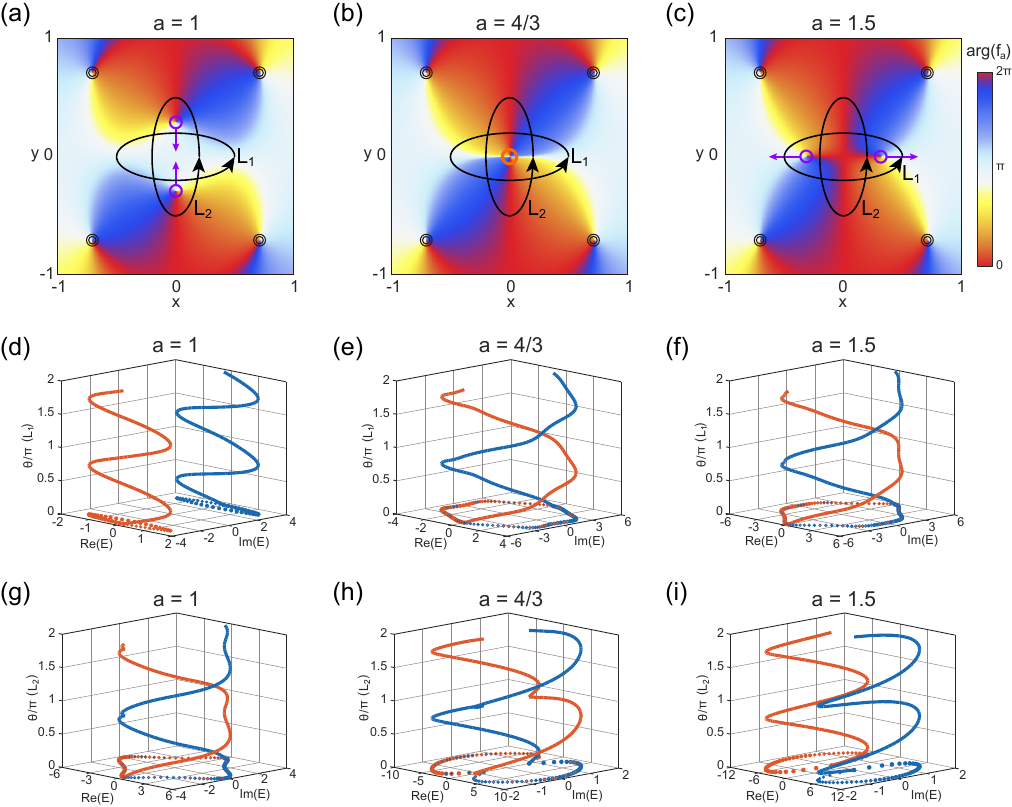}
\caption{\textcolor{red}{(a-c) Zoomed-in Figs.~\ref{K41_PT}(a-c) around original point. Two oriented curved are given by $L_1: (r_1cos(\theta), r_2sin(\theta)), \theta:0 \rightarrow 2\pi$ with $(r_1,r_2)=(0.5, 0.2)$; $L_2: (r_1cos(\theta), r_2sin(\theta)), \theta:0 \rightarrow 2\pi$ with $(r_1,r_2)=(0.2, 0.5)$. Color scale: $arg(f_a)$ in $xy$-plane. Circles: vortices in $arg(f)$ by using standard stereographic mapping Eq.~(\ref{stereographic_coordinates}), which correspondence to exceptional points (EPs). The arrows indicate the moving of EPs. The value of scaling parameter $a$ is given at the top of each panel. (d-f) Evolution of complex energy braiding in the $(Re(E),Im(E),\theta)$-space along $L_1$ is varied as $a$. (g-i) Evolution of complex energy braiding in the $(Re(E),Im(E),\theta)$-space along $L_2$ is varied as $a$.} }
\label{CE_braid}
\end{center}
\end{figure*}
%%%%%%%%%%%%%%%%%%%%%%%%%%%%%%%%%%%
%%%%%%%%%%%%%%%%%%%%%%%%%%%%%%%%%%%

\section{Topological transitions: untie exceptional konts} \label{untie_TPT}

In our Non-Hermitian systems, we implement a controlled unknotting via a single, continuously tunable parameter $a$. 
\textcolor{red}{To classify the local evolution of exceptional lines, we consider three oriented links $L_+$, $L_-$ and $L_0$ [see Figs.~S2(a-c)], which are identical except in a small neighborhood where they differ as follows: $L_+$ and $L_-$ contain a positive and a negative crossing, respectively, while $L_0$ is obtained by performing the oriented smoothing of the crossing. We identify two distinct types of local topological transitions~\cite{Jones_zensen}. Type-I transition is a crossing change mediated by an EC, corresponding to the conversion between positive and negative crossings ($L_+ \leftrightarrow L_-$ ) or the inversion of braid words ($\sigma_i \leftrightarrow \sigma_i^{-1}$). Type-II transition involves the breaking and recombination of exceptional lines, facilitating the creation or removal of crossings ($L_0 \leftrightarrow L_\pm$) akin to the braid relations ($\sigma_0 \leftrightarrow \sigma_i^{\pm}$). It is crucial to note that} the physical pathway induced by increasing $a$ does not strictly correspond to the mathematical \textit{unknotting number}. The unknotting number is defined as the minimum number of crossing changes to convert a knot into the unknot, \textcolor{red}{where only the type-I transition is permitted. Our system utilizes both type-I and type-II processes to achieve the same topological simplification.} We demonstrate this principle through the deliberate untying of the Figure-eight exceptional knot.

The function $F_a$ is defined on $\mathbb{C}^2$, but we are interested in its restriction to the unit three-sphere $\mathcal{S}^3 = \left\{ (u, v)\in\mathbb{C}^2: |u|^2+|v|^2=1 \right\}$. For a better visualization of exceptional knots, in the following analysis, we will adopt the standard stereographic coordinates to map $\mathcal{S}^3$ to $3$D Euclidean space $\mathbb{R}^3$
\begin{equation}
    u = \frac{x^2 + y^2 + z^2 - 1 + 2iz}{x^2 + y^2 + z^2 + 1 },v = \frac{2(x + iy)}{x^2 + y^2 + z^2 + 1 }. \label{stereographic_coordinates}
\end{equation}

The initial exceptional Figure-eight knot at $a = 1$ is obtained from Eq.~(\ref{Fa_L3n2}) with $n=2$, shown in Fig.~\ref{K41_PT}(a), fully contained within a compact region of the Brillouin zone. Upon increasing the parameter to the critical value $a = 4/3$ [Fig.~\ref{K41_PT}(b)], a significant reconfiguration occurs. The inner segments of the knot, which locally resemble hyperbolic branches, meet together at the $\Gamma$ point (the origin of the Brillouin zone), forming a flat ribbon-like structure identified as a EC. Simultaneously, the outer loops of the original knot break, transforming into an open EC that extend outward. A further increase to $a=3/2$, corresponding to Fig.~\ref{K41_PT}(c), completes the unknotting process through \textcolor{red}{both type-I and type-II transitions}. The central EC splits apart, resolving the previous entanglement of the inner branches. Simultaneously, the open EC from the previous stage reconnect with each other, contracting and merging to form a single, $3$D coiled unknotted ER.

The morphological evolutions of exceptional knots described above are directly linked to the underlying movements and reconnection of the associated EPs, as revealed by the phase of the function $f$ in the $xy$-plane, shown in Figs.~\ref{K41_PT}(d-f). The vortices in $arg(f)$, marked by circles, correspond to the EPs whose trajectories trace exceptional knots in $3$D momentum space. \textcolor{red}{The associated charge with each EP is quantified as~\cite{wang2025non}
\begin{equation}
    Q(L) = \frac{1}{2\pi}\oint_{L(\beta)} \nabla arg[f(\beta)] \cdot d\beta, \label{chargeQ}
\end{equation}
where $L(\beta)$ is a closed curve encircling the EP. An EP acts as a source or sink of Berry curvature, and its charge determines the chiral behavior of eigenstates upon encircling the singularity. And the charge is positive(negative) when the band-1($E_1$) crosses the band-2($E_2$) from above(below) in complex-energy braiding on the trajectory $L$. Specially, $Q=+1(-1)$ corresponds to a complex-energy braiding $\sigma_1$($\sigma_1^{-1}$) [see fig.~S6(j)], and $Q=+2$ to a double $\sigma_1^2$ [see figs.~\ref{CE_braid}(e) and (h)].} 

Around $\Gamma$ point, four EPs (marked as black circles) \textcolor{red}{with charges $Q=-1$} remain stationary in all cases in Figs.~\ref{K41_PT}(d-f) during the process. While two other pairs of EPs \textcolor{red}{with charges $Q=+1$}, indicated by turquoise and purple circles, exhibit distinct motions as $a$ increases, as indicated by arrows. Firstly, the purple EPs are moving slowly towards the $\Gamma$ point along the $y$-axis, while the turquoise EPs are moving rapidly away from $\Gamma$ point along the $x$-axis. As $a$ increases to $4/3$ [Fig.~\ref{K41_PT}(e)], the purple vortices meet at the $\Gamma$ point, coinciding with the formation of the central EC in Fig.~\ref{K41_PT}(b). Simultaneously, the vanishment of turquoise EPs pair at infinite coincides with the breaking of the knot. Subsequently, as $a$ further increases to $3/2$, the subsequent rearrangement and separation of these EP pairs facilitate the splitting of the EC and the reconnection of the open exceptional arcs into the final unknotted ER in Fig.~\ref{K41_PT}(c).

\textcolor{red}{To further characterize the type-I topological transition, we analyze the spectral braiding topology along two oriented loops $L_{1,2}$ defined in Figs.~\ref{CE_braid}(a-c). The type-I transition can be characterized by a pair of winding numbers $( Q(L_1),Q(L_2) )$. The associated topological charges are evaluated through Eq.~(\ref{chargeQ}), where the sign and index of the complex-energy braiding encode the net number of enclosed EPs. For $a=1$, corresponding to the initial Figure-eight exceptional knot phase, the charges are $(0,2)$ [Figs.~\ref{CE_braid}(d) and (g)]. At the critical point $a=4/3$. where the EC emerges and the crossing inversion occurs, the braiding topology becomes $(2,2)$ [Figs.~\ref{CE_braid}(e) and (h)]. After the transition is completed at $a=1.5$, the final unknotted ER phase is characterized by (2, 0)[Figs.~\ref{CE_braid}(f) and (i)]. Therefore, the type-I transition can be understood as a transfer of spectral winding between the two oriented loops, accompanied by the reconnection and redistribution of EP pairs in momentum space. Similarly, the type-II transition can be characterized by a pair of winding numbers $( Q(L_3),Q(L_4) )$, as illustrated in Fig.~S6.}

\section{Summary and Discussions} \label{Summary}

\textcolor{red}{We have developed a universal and constructive framework for realizing exceptional knots and links with arbitrary braiding topology in minimal two-band non-Hermitian systems. 
The universality claimed here refers to the constructive realization of arbitrary braid topologies once a braid representation is given, rather than to the separate knot-theoretic problem of determining minimal braid representatives.} 
By establishing a direct mapping from braid representations to semiholomorphic polynomials and tight-binding Hamiltonians, our approach unifies previously known constructions of knotted exceptional structures~\cite{knot_2019PRB, knot_2021CP, knot_2021PRL, link_2018PRA, link_2019PRB, link_2020PRB, link_2020PRA} within a single theoretical framework and demonstrates that arbitrary knot and link topologies can be systematically engineered without relying on symmetry protection or fine-tuning. In contrast to Hermitian nodal structures, whose stability and topology are strongly constrained by crystalline or internal symmetries~\cite{2011PRB_lines, 2017PRB_links, 2018PRB_links, 2016Nature_chains, 2018NP_chains}, exceptional lines arise generically in three-dimensional non-Hermitian systems as intersections of real surfaces in momentum space. This intrinsic robustness \textcolor{red}{enables the realization of intricate knotted and linked configurations that naturally encode global topological information}.

Beyond static constructions, we further demonstrated the continuous untying of exceptional knots through a single tunable parameter. 
\textcolor{red}{The resulting topological transitions proceed via the motion and reconnection of EPs, accompanied by the transient formation of exceptional chains.} 
Unlike conventional Hermitian topological phase transitions governed by isolated gap closings, \textcolor{red}{these processes involve global reconnections of the degeneracy manifold and enable deterministic transformations between distinct knot topologies}. This establishes a controllable route toward manipulating knot complexity in momentum space and suggests that knotted exceptional structures may serve as tunable topological resources in non-Hermitian systems.

The generality of our framework opens several promising directions for future study. Extensions to multiband non-Hermitian systems may support richer knotted structures associated with higher-order exceptional degeneracies and \textcolor{red}{non-Abelian} band permutations. \textcolor{red}{It is worth noting that while the two-band models considered here feature Abelian exceptional points characterized by integer winding numbers, recent studies have revealed that multi-band non-Hermitian systems can host non-Abelian braiding invariants where single EPs carry non-trivial topological charges~\cite{2021PRB_homotopy,2022PRB_FSH,PRR_2023BEJ,2026PRL_singleEP}. The knotting and untying transitions discussed in this work pertain to the geometric arrangement of standard Abelian EPs in 3D momentum space.} It will also be interesting to explore the interplay between knotted exceptional topology and additional symmetries, including crystalline and PT symmetries, as well as their dynamical consequences in adiabatic evolution, nonreciprocal transport, and nonlinear wave phenomena. More broadly, the braid-based polynomial construction introduced here may provide a versatile platform for engineering knotted structures across diverse physical settings, including photonics~\cite{wang2025topological, wang2026topological}, acoustics~\cite{acoustic_2020NC}, cold atoms~\cite{link_2020PRA}, liquid crystals~\cite{2017PRX, hall2025fusion}, and topological field theories~\cite{amari20183}.

\section*{Acknowledgments}
This work was supported by the National Key R\&D Program of China (2022YFA1404400), National Natural Science Foundation of China under grant Nos. 12125504 and 52306109, the CAS Pioneer Hundred Talents Program, Priority Academic Program Development (PAPD) of Jiangsu Higher Education Institutions, the Doctoral Support Program for Young Talents of the China Association for Science and Technology. We thank Han Zeng and Ruo-Yang Zhang for useful discussion.

\bibliography{ref}

\end{document}